\documentclass[12pt]{article}
\usepackage{amsmath}
\usepackage{amsfonts}
\usepackage{amssymb}
\usepackage{srcltx}
\usepackage{arydshln}

\newtheorem{theorem}{Theorem}

\newtheorem{definition}[theorem]{Definition}

\newtheorem{remark}[theorem]{Remark}

\begin{document}
\hfill\today\\
\noindent
{\Large {\bf Polynomial tau-functions for the multi-component KP hierarchy}}
\vskip 9 mm
\begin{center}
\begin{minipage}[t]{70mm}
{\bf Victor G. Kac}\\
\\
Department of Mathematics,\\
Massachusetts Institute of Technology,\\
Cambridge, Massachusetts 02139, U.S.A\\
e-mail: kac@math.mit.edu\\
\end{minipage}\qquad
\begin{minipage}[t]{70mm}
{\bf Johan W. van de Leur}\\
\\
Mathematical Institute,\\
Utrecht University,\\
P.O. Box 80010, 3508 TA Utrecht,\\
The Netherlands\\
e-mail: J.W.vandeLeur@uu.nl
\end{minipage}
\end{center}
\begin{abstract}
In a previous paper 
we constructed all polynomial tau-functions of the 1-component KP hierarchy, namely, we showed that any such tau-function is obtained from a Schur polynomial $s_\lambda(t)$ by certain shifts of arguments. In the present paper we give a simpler proof of this result,   using the (1-component) boson-fermion correspondence. Moreover, we show that this approach can be applied to the $s$-component KP hierarchy, using the $s$-component boson-fermion correspondence, finding thereby all its polynomial tau-functions.
We also find all polynomial tau-functions for the reduction of the $s$-component KP hierarchy, associated to any 
partition consisting of $s$ positive parts.\\
\
\\{\it Keywords:} KP hierarchy, multi-component KP, tau-functions, Schur polynomials\\
\
\\
{\it Math. Subject Classification (2010):} 14M15, 17B10, 17B67, 20G44, 22E70, 35Q53
\end{abstract}
\section{Introduction}
\label{S1}
In his paper \cite{S} M. Sato introduced the (1-component) KP hierarchy of evolution equations of Lax type on the pseudo-differential operator 
\begin{equation}
\label{E1}
L=\partial+u_1(t)\partial^{-1}+u_2(t)\partial^{-2}+\cdots,
\end{equation}
where $t=(t_1,t_2,\ldots)$ and $\partial=\frac{\partial}{\partial t_1}$,  and the corresponding tau-function $\tau(t)$. Moreover for any positive integer $s$, he also introduced the $s$-component  KP hierarchy  on the $s\times s$- matrix pseudo-differential operator $L$ of the form (\ref{E1}), where $u_i(t)$ are $s\times s$-matrices  and $t=(t_j^{(k)}
\, | \, j=1,2,\ldots; k=1,2,\ldots, s)$ and $\partial =\frac{\partial}{\partial t_1^{(1)}}+\cdots+\frac{\partial}{\partial t_1^{(s)}}$, along with certain subsidiary equations (which are absent for $s=1$).

In the subsequent paper \cite{DJKM} the tau-function was introduced for the $s$-component KP hierarchy, generalizing the one by Sato for $s=1$. The theory was further developed in \cite{KL}.
Since any solution of the $s$-component KP hierarchy is explicitly expressed in terms of the tau-function, it is important to construct the latter.

In our paper \cite{KvdL-MKP} we constructed all polynomial tau-functions of the 1-component KP hierarchy, namely, we showed that any such tau-function is obtained from a Schur polynomial $s_\lambda(t)$ by certain shifts of arguments. In the present paper we give a simpler proof of this result (Theorem \ref{T7}),   using the (1-component) boson-fermion correspondence. Moreover, we show that this approach can be applied to the $s$-component KP hierarchy, using the $s$-component boson-fermion correspondence, finding thereby all its polynomial tau-functions (Theorem \ref{mT}).

In paper \cite{KL} we studied the reduction of the $s$-component KP hierarchy associated to any $s$-part partition $\lambda=\{n_1\ge n_2\ge\cdots\ge n_s>0\}$. A special case is the Gelfand-Dickey $n$-th KdV, associated to the 1-part  partition $\lambda=\{ n\}$, for which we found in \cite{KvdL-MKP} all polynomial tau-functions. In the present paper we reprove this result, using boson-fermion correspondence (Theorem \ref{TKdV}). We use the same method to find all polynomial tau-functions for the $\lambda$-reduced $s$-component KP hierarchy (Theorem \ref{mmT}). In conclusion of the paper we consider, as the simplest example beyond the 1-part partition, the 2-part partition $1+1$. This produces the AKNS ($=$ non-linear Schr\"odinger hierarchy) see \cite{KL}, and we find all its polynomial tau-functions.

\section{The  fermionic formulation of the KP hierarchy}
\label{S2}

The  group
$GL_{\infty}$, consisting of all complex matrices  $G= (g_{ij})_{i,j \in \mathbb{Z}}$ which are 
 invertible and all but a finite number of $g_{ij} -
\delta_{ij}$ are $0$,
acts  on the vector space 
$\mathbb{C}^{\infty} = \bigoplus_{j \in \mathbb{Z}} {\Bbb
C} e_{j}$, via the  formula
$E_{ij} (e_{k}) = \delta_{jk} e_{i}$.

The semi-infinite wedge space (see e.g. \cite{KP}, \cite {KL}) $F =
\Lambda^{\frac{1}{2}\infty} \mathbb{C}^{\infty}$ is the vector space
with a basis consisting of all semi-infinite monomials of the form
$e_{i_{1}} \wedge e_{i_{2}} \wedge e_{i_{3}} \ldots$, where $i_{1} >
i_{2} > i_{3} > \ldots$ and $i_{\ell +1} = i_{\ell} -1$ for $\ell >>
0$.  One defines the  representation $R$ of $GL_{\infty}$  on $F$ by
$$
R(G) (e_{i_{1}} \wedge e_{i_{2}} \wedge e_{i_{3}} \wedge \cdots) = G
e_{i_{1}} \wedge G e_{i_{2}} \wedge Ge_{i_{3}} \wedge \cdots , 
$$
and apply linearity and anticommutativity of  the wedge product $\wedge$.

The corresponding representation $r$ of the Lie algebra
$g\ell_{\infty}$
of $GL_\infty$ can be described
in  terms  wedging and contracting
operators $\psi^{+}_{j}$ and $\psi^{-}_{j}\ \ (j \in \mathbb{Z} +
\frac{1}{2})$ on $F$:
$$\begin{aligned}
\psi^{+}_{j} (e_{i_{1}} \wedge e_{i_{2}} \wedge \cdots )&=  
e_{-j+\frac12}\wedge e_{i_{1}} \wedge e_{i_{2}} \cdots, \\
\psi^{-}_{j} (e_{i_{1}} \wedge e_{i_{2}} \wedge \cdots ) &= \begin{cases} 0
&\text{if}\ j-\frac12 \neq i_{s}\ \text{for all}\ s \\
(-1)^{s+1} e_{i_{1}} \wedge  \cdots \wedge
e_{i_{s-1}} \wedge e_{i_{s+1}} \wedge \cdots &\text{if}\ j = i_{s}-\frac12.
\end{cases}
\end{aligned}
$$
These operators satisfy the relations of a Clifford algebra, which we denote by ${\cal C}\ell$:
$(i,j \in \mathbb{Z}+\frac{1}{2}, \lambda ,\mu = +,-)$:
$$\psi^{\lambda}_{i} \psi^{\mu}_{j} + \psi^{\mu}_{j}
\psi^{\lambda}_{i} = \delta_{\lambda ,-\mu} \delta_{i,-j}. $$
Let  $(k \in \mathbb{Z})$
\begin{equation}
\label{vac}
|k\rangle = e_{k } \wedge e_{k-1 } \wedge
e_{k-2} \wedge \cdots ,
\end{equation} 
then $F$ is an irreducible ${\cal C}\ell$-module, such that
$$\psi^{\pm}_{j} |0\rangle = 0 \ \text{for}\ j > 0 . $$
The representation $r$ of $g\ell_\infty$ in $F$, corresponding to the representation $R$ of $GL_\infty$, is given by the
 formula
$r(E_{ij}) = \psi^{+}_{-i+\frac12} \psi^{-}_{j-\frac12}. $
Define the {\it charge decomposition}
$$F = \bigoplus_{k \in \mathbb{Z}} F^{(k)}, \quad \text{
where }
\text{charge}(|k\rangle ) = k\ \text{and charge} (\psi^{\pm}_{j}) =
\pm 1. $$
The space $F^{(k)}$ is an irreducible highest weight
$g\ell_{\infty}$-module, with highest weight vector
$|k\rangle$:
$$
r(E_{ij})|k\rangle = 0 \ \text{for}\ i < j,\quad
r(E_{ii})|k\rangle = 0\  (\text{resp.}\ = |k\rangle ) \ \text{if}\ i > k\
(\text{resp. if}\ i \le k).
$$
Let
$${\cal O}_k
= R(GL_{\infty})|k\rangle \subset F^{(k)}$$ be the $GL_{\infty}$-orbit
of the highest weight vector $|k\rangle$.

\begin{theorem}{\rm  (\cite{KP}, Theorem 5.1)}
Let  $0\not =g_k\in F^{(k)}$.
Then $g_k\in  {\cal O}_k$ if and only if
\begin{equation}\
\label{KP}
\sum_{i\in\mathbb{Z}+
\frac{1}{2}} \psi_i^+ g_k\otimes \psi_{-i}^- g_k =0\, .
\end{equation}
\end{theorem}

Equation (\ref{KP}) is called the  KP hierarchy in the fermionic picture.

Now recall that the orbit
${\cal O}_k$ of $\mathbb{C} |k\rangle\in F^{(k)}$  of the group $GL_\infty$, for a fixed $k\in\mathbb{Z}$ can be decomposed as follows. 
Let $P$ be the stabilizer of  $\mathbb{C} |k\rangle$,
let $W$ be the subgroup of permutations of basis vectors of $\mathbb{C}^\infty$ and let $W_k$ be its subgroup, consisting of permutations, permuting vectors with indices less or equal to $k$ between themselves. Then one has the Bruhat decomposition:
\[ GL_\infty = \bigcup_{w\in W/W_k} UwP \,\,\,(\hbox{disjoint union}).\]
Applying this to $\mathbb{C} |k\rangle$, we obtain that the projectivised orbit
$\mathbb{P} {\cal O}_k$ is a disjoint union of Schubert cells $C_w =U w\cdot |k\rangle$, $w\in W/W_k$.
It is well known (see, e.g. \cite{KP}) that each $w\cdot |k\rangle$ corresponds to a partition $\lambda=\lambda (w)$, and the corresponding Schubert cell $C_\lambda =Uw(\lambda)\cdot |k\rangle $, where $U$ is the subgroup of $GL_\infty$, consisting of upper triangular
matrices with $1$'s on the diagonal, is an affine algebraic variety isomorphic to $\mathbb{C}^{|\lambda|}$.

\section{The  bosonic formulation of KP}
\label{S3}
Define the fermionic fields by 
$
\psi^\pm(z)=\sum_{i\in\mathbb{Z}+\frac12}\psi_i^\pm z^{-i-\frac12}
$
and the bosonic field
$\alpha(z)=\sum_{n\in\mathbb{Z}}\alpha_n z^{-n-1}=:\psi^+(z)\psi^-(z):$.
Then there exists a unique vector space isomorphism, called the boson-fermion correspondence,
$ \sigma:F \to B=\mathbb{C}[q,q^{-1}]\otimes \mathbb{C}[t_1,t_2,\ldots]$ 
such that $\sigma (|k\rangle)=q^k$, $\sigma \alpha_n\sigma^{-1}=\frac{\partial}{\partial t_n}$, 
$\sigma \alpha_{-n}\sigma^{-1}=n t_n$, for $n>0$ and  $\sigma \alpha_0\sigma^{-1}=q\frac{\partial}{\partial q}$. Moreover, one has
\begin{equation}
\label{2}
\sigma \psi^\pm(z)\sigma^{-1}= q^{\pm 1}z ^{\pm q\frac{\partial}{\partial q}}\exp\left(\pm \sum_{k=1}^\infty t_kz^k\right)
\exp\left(\mp \sum_{k=1}^\infty \frac{\partial}{\partial  t_k}\frac{ z^{-k}}{k}\right).
\end{equation}
Note that $Q=\sigma^{-1} q\sigma$ is the following operator on $F$:
\begin{equation}
\label{Q}
Q|k\rangle =|k+1\rangle\quad\mbox{and}\quad Q\psi^{\pm}_i=\psi^{\pm}_{i\mp 1}Q\, .
\end{equation}
For $g_k\in{\cal O}_k\cup\{ 0\}$  we write: $\sigma (g_k)= \tau_k(t) q^k$, where $t=(t_1,t_2,\ldots)$. Such a $\tau_k$ is called a tau-function. 
It is well known see e.g. \cite{KP} that $\tau_k(t)$ is equal to the coefficient of $|k\rangle$ in $\exp\left(\sum_{i=1}^\infty t_i\alpha_i \right) g_k$.
Under the isomorphism $\sigma$ we can rewrite (\ref{KP}), using (\ref{2}), to obtain a Hirota bilinear identity for tau-functions, see e.g. \cite{DJKM} or \cite{KP}.
Let 
$[z]=(z,\frac{z^2}2,\frac{z^3}3,\ldots)$, $y=(y_1,y_2,\ldots)$, and ${\rm Res}\,  \sum_i f_iz^i dz=f_{-1}$, then 
\begin{equation}
\label{KP2}
{\rm Res}\,   \tau_k(t-[z^{-1}])\tau_k(y+[z^{-1}]) \exp\left(\sum_{i=1}^\infty (t_i-y_i)z^i\right)dz=0.
\end{equation}

\section{Polynomial solutions of KP}
\label{S7}
Introduce  the  elementary Schur polynomials by
\begin{equation} 
\label{elSchur}\exp\left(\sum_{i=1}^\infty t_iz^i\right)=\sum_{j=0}^\infty s_j(t)z^j\, , \end{equation} 
then in \cite{KvdL-MKP} we proved the following :
\begin{theorem}
\label{T7}
All polynomial tau-functions of the KP hierarchy are, up to a constant factor, of the form
\begin{equation}
\label{Pt}
\tau_{\lambda_1,\lambda_2,\ldots,\lambda_m}(t;c_1,c_2,\ldots ,c_m)=\det \left( s_{\lambda_j+i-j}(t_1+c_{1,j},t_2+c_{2j},t_3+c_{3j},\ldots)\right)_{1\le i,j\le m}\, ,
\end{equation}
where $\lambda=(\lambda_1,\lambda_2, \ldots,\lambda_m)$ is a partition and
$c_j=(c_{1j},c_{2j},\ldots) \in\mathbb{C}^{\lambda_j+m-j}$ are arbitrary.\\
\end{theorem}
Here we will give a simpler proof of this theorem.
\\
\
\\
{\bf Proof.} Fix an integer $k$. Then the  element $w(\lambda)|k\rangle$ in the Schubert cell $C_\lambda$ of $O_k$, corresponding to the partition $\lambda=(\lambda_1,\lambda_2, \ldots, \lambda_m)$, is the element
\begin{equation}
\label{lambda}
e_{\lambda_1+k}\wedge e_{\lambda_2+k-1}\wedge\cdots\wedge e_{\lambda_m+k-m+1}\wedge e_{k-m}\wedge e_{k-m-1}\wedge\cdots\ .
\end{equation}
If we let an element $u\in U$, an upper triangular matrix with $1$'s on the diagonal, act on this element we obtain the element 
\begin{equation}
\label{lambda2}
f_{1}\wedge f_{2}\wedge\cdots\wedge f_{m}\wedge e_{k-m}\wedge e_{k-m-1}\wedge\cdots\ ,
\end{equation}
where
\begin{equation}
\label{lambda3}
f_{j}= e_{\lambda_j+k-j+1}+\sum _{i=k-m+1}^{\lambda_j+k-j}a_{ij}e_i\, .
\end{equation}
Note that we can use Gauss elimination with which we can eliminate certain $a_{ij}$, viz., we may assume that $a_{\lambda_i+k-i+1,j}=0$ for all $i>j$.
This means that we can put $1+2+\cdots +m-1=\frac12(m-1)m$ constants to zero, all others are arbitrary. Counting all coefficients which are arbitrary, we find 
\[
-\frac12(m-1)m+\sum_{j=1}^m (\lambda_j+k-j)-(k-m)=\frac12(m-1)m+\sum_{j=1}^m \lambda_j+\sum_{j=1}^m (m-j)=\sum_{j=1}^m \lambda_j
\, .
\]
Hence, we find indeed that the dimension of the Schubert cell $C_\lambda$ is equal to $|\lambda |$.
\\
Assume from now on that all $a_{ij}$ of (\ref{lambda3}) are again arbitrary,  then we want to calculate the image $\tau^{(k)}_\lambda(t)$ of the element (\ref{lambda2}) under the isomorphism $\sigma$. Recall from e.g. \cite{KP} that $\tau^{(k)}_\lambda(t)$ is the coefficient of $|k \rangle $ in 
\begin{equation}
\label{lambda4}
\exp\left(\sum_{i=1}^\infty t_i\alpha_i \right)f_{1}\wedge\cdots\wedge f_{m}\wedge e_{k-m}\wedge e_{k-m-1}\wedge\cdots=
f_{1}(t)\wedge \cdots\wedge f_{m}(t)\wedge e_{k-m}\wedge e_{k-m-1}\wedge\cdots ,
\end{equation}
where, since $\alpha_i$ is a derivation of the exterior product, such that $\alpha_i(e_j)=e_{j-i}$, we have 
\begin{equation}
\label{lambda5}
\begin{aligned}
f_{j}(t)=&\exp\left(\sum_{i=1}^\infty t_i\alpha_i \right)f_{j}\\
=&\sum_{\ell=0}^\infty s_\ell(t) \left(e_{\lambda_j-j+1+k-\ell} +\sum_{i=k-m+1}^{\lambda_j-j+k} a_{ij} e_{i-\ell}\right)\\
=&\sum_{p=0}^\infty \left(
s_p(t)+\sum_{i=1}^p a_{\lambda_j-j+1+k-i,j}s_{p-i}(t)
\right)e_{\lambda_j-j+1+k-p}
\, .
\end{aligned}
\end{equation}
Since $e_{k-m}, e_{k-m-1},\ldots$ appear in (\ref{lambda2}), we can replace $f_j(t)$ by the  element (\ref{lambda5})  where all $e_s$ with $s\le k-m$ are removed. In other words we may assume that 
\begin{equation}
\label{lambda6}
f_{j}(t)=
\sum_{p=0}^{\lambda_j-j+m} \left(
s_p(t)+\sum_{i=1}^p a_{\lambda_j-j+1+k-i,j}s_{p-i}(t)
\right)e_{\lambda_j-j+1+k-p}
\, .
\end{equation}
Since, by (\ref{elSchur}),  $\frac{\partial s_a(t)}{\partial t_1}=s_{a-1}(t)$, equation (\ref{lambda6}) can be rewritten as
\begin{equation}
\label{lambda66}
f_{j}(t)=
\sum_{r=0}^{\lambda_j-j+m} 
\frac{\partial^r \left( s_{\lambda_j-j+m}(t)+\sum_{i=1}^{\lambda_j-j+m} a_{\lambda_j-j+1+k-i,j}s_{{\lambda_j-j+m}-i}(t)\right)}{\partial t_1^r}
e_{k-m+1+r}
\, .
\end{equation}
Next, it follows from (\ref{elSchur})  that 
\begin{equation}
\label{eq15}
s_j(t+c)=\sum_{i=0}^j s_{j-i}(c)s_i(t)\, .
\end{equation}
Note also that the map $(s_1, \ldots , s_n):\mathbb{C}^n\to \mathbb{C}^n$ is surjective. Hence there exist constants $c_j=(c_{1j},c_{2j},\dots)$, such that 
\begin{equation}
\label{aconst}
a_{k-p+1,j}=s_{\lambda_j-j+p}(c_j)\, .
\end{equation} 
It follows from (\ref{eq15}) that 
\[
s_{\lambda_j+m-j}(t+c_j)=s_{\lambda_j+m-j}(t)+\sum_{\ell=0}^{\lambda_j+m-j-1} s_{\lambda_j+m-j-\ell}(c_j)s_\ell(t)\\
\]
Thus, using (\ref{aconst}), we obtain from (\ref{lambda66})
\begin{equation}
\label{lambda7}
f_{j}(t)=\sum_{r=0}^{\lambda_j-j+m}\frac{\partial^r s_{\lambda_j-j+m}(t+c_j)}{\partial t_1^r}e_{k-m+r+1}=\sum_{r=0}^{\lambda_j-j+m}s_{\lambda_j-j+m-r}(t+c_j)e_{k-m+r+1}\, .
\end{equation}
Hence the coefficient of $|k\rangle$ in (\ref{lambda4}) is equal to (cf. \cite{KP})
$$
\det \left( s_{\lambda_j+i-j}(t_1+c_{1,j},t_2+c_{2j},t_3+c_{3j},\ldots)\right)_{1\le i,j\le m}\, ,
$$
which is the desired result.\hfill$\square$
\
\\
\begin{remark}
Different $c_{ij}$'s can give the same polynomial solutions in $C_\lambda$.
From  the proof of  Theorem \ref{T7}, it is clear how to obtain a one to one correspondence in terms of the constants $a_{ij}$, viz. one has to assume that all
$a_{\lambda_i+k-i+1,j}=0$, for all $1\le j<i\le m$. Using (\ref{aconst}) this means that $s_{\lambda_j-j-\lambda_i+i}(c_j)=0$, for all $1\le j<i\le m$,
or stated differently,
\begin{equation}
\label{restriction}
c_{\lambda_j-j-\lambda_i+i,j}=-s_{\lambda_j-j-\lambda_i+i}(c_{1j}, c_{2j},\ldots, c_{\lambda_j-j-\lambda_i+i-1,j},0)\quad\mbox{ for all } 1\le j<i\le m.
\end{equation}
This means that for fixed $j$ we find constants $c_{ij}$'s for $1\le i\le \lambda_j-j+m$, of which  we can eliminate all $c_{\lambda_j-j-\lambda_i+i,j}$, with $j<i\le m$, by using  formula (\ref{restriction}) recursively. 
Note that for fixed $j$ there are $\lambda_j-j+m$ constants $c_{ij}$'s of which $m-j$ can be eliminated. Thus we have $\lambda_j$ constants $c_{ij}$. Hence in total, we have $|\lambda |=\lambda_1+\lambda_2+\cdots+\lambda_m$ constants. This agrees with the fact that $\dim C_\lambda=|\lambda |$.
\end{remark}

\section{The multicomponent KP}
In this section we  introduce the $s$-component KP, where $s$ is a positive integer. 
In \cite{KL} we introduced a relabeling of the $e_i$ to obtain the multicomponent KP. One approach would be to use this relabeling and describe all vectors (\ref{lambda3}) of the semi-infinite wedges (\ref{lambda2}) in terms of these new relabeled $e_i$'s and determine its corresponding  tau-functions, using the multicomponent bosonization. But in that case the solution depends on the relabeling and one does not obtain all possible solutions, but one obtains all solutions related to that relabeling. To obtain all solutions one has to calculate all solutions corresponding to all possible relabelings. Instead, we will introduce a new basis of $\mathbb{C}^\infty$, viz., $e_i^{(a)}$ where $1\le a\le s$ and $i\in\mathbb{Z}$. We assume that $|0\rangle$ is the semi-infinite wedge vector consisting of all $e_i^{(a)}$ with $1\le a\le s$ and $i\le 0$. We introduce creation (+) and annihilation (-) operators,
$\psi_i^{\pm(a)}$, with, as before, $i\in \frac12+\mathbb{Z}$, such that ($\lambda, \mu=+$ or $-$)
$$\psi^{\lambda(a)}_{i} \psi^{\mu(b)}_{j} + \psi^{\mu(b)}_{j}
\psi^{\lambda(a)}_{i} = \delta_{\lambda ,-\mu}\delta_{ab} \delta_{i,-j}. $$
and
\[
\psi^{\pm (a)}_i|0\rangle=0,\quad \mbox{for }i>0\ \mbox{and }\psi_i^{+(a)}|0\rangle= e^{(a)}_{-i+\frac12}\wedge |0\rangle\, .
\]
We have the same charge decomposition, where now the charge  of $|0\rangle$ is again 0 and the charge of $\psi_i^{\pm(a)}$ is $\pm 1$.
The disadvantage of this approach is that we cannot describe $\sigma(|k\rangle)$, for $k\not = 0$, explicitly.

The KP hierarchy (\ref{KP}), thus turns into the $s$-component KP in the fermionic picture:
\begin{equation}
\label{mKP}
\sum_{a=1}^s\sum_{i\in\mathbb{Z}+
\frac{1}{2}} \psi_i^{+(a)} g_k\otimes \psi_{-i}^{-(a)} g_k =0,\quad g_k\in F^{(k)}\, .
\end{equation}

Define the fermionic fields by 
$
\psi^{\pm(a)}(z)=\sum_{i\in\mathbb{Z}+\frac12}\psi_i^{\pm(a)} z^{-i-\frac12}
$
and the bosonic fields
$\alpha^{(a)}(z)=\sum_{n\in\mathbb{Z}}\alpha_n^{(a)} z^{-n-1}=:\psi^{+(a)}(z)\psi^{-(a)}(z):$, where $a=1,2,\ldots, s$. 
As in the 1-component case we need additional operators $Q_a$, $a=1,2,\ldots, s$, to define the boson-fermion correspondence. They are uniquely defined by 
\begin{equation}
\label{Qa}
Q_a|0\rangle =\psi_{-\frac12}^{+(a)}|0\rangle\quad\mbox{and}\quad Q_a\psi^{\pm(b)}_i=(-1)^{1-\delta_{ab}}\psi^{\pm(b)}_{i\mp \delta_{ab}}Q_a\, .
\end{equation}
It is straightforward to check that $Q_a$ and $Q_b$ anti-commute if $a\ne b$ and that 
\[
Q_a\alpha_j^{(b)}=\alpha_j^{(b)}Q_a-\delta_{ab}\delta_{j0}Q_a,\quad 1\le a,b\le s\, .
\]
Then as before 
 there exists a unique vector space isomorphism
$ \sigma:F \to B=\mathbb{C}[q_a,q_a^{-1};1\le a\le s]\otimes \mathbb{C}[t_1^{(a)},t_2^{(a)},\ldots;1\le a\le s]$ 
such that $\sigma (|0\rangle)=1$, $\sigma \alpha_n^{(a)}\sigma^{-1}=\frac{\partial}{\partial t_n^{(a)}}$, 
$\sigma \alpha_{-n}^{(a)}\sigma^{-1}=n t_n^{(a)}$, for $n>0$ and  $\sigma \alpha_0^{(a)}\sigma^{-1}=q_a\frac{\partial}{\partial q_a}$. 
Then $\sigma Q_a \sigma^{-1}=\epsilon_aq_a$, 
where $\epsilon_a q_b=(-1)^{1-\delta_{ab}}q_b\epsilon_a$ and $\epsilon_a
|0\rangle=|0\rangle$.
Moreover, one has (cf. \cite{KL})
\begin{equation}
\label{2m}
\sigma \psi^{\pm(a)}(z)\sigma^{-1}= \epsilon_aq_a^{\pm 1}(z) ^{\pm q_a\frac{\partial}{\partial q_a}}\exp\left(\pm \sum_{k=1}^\infty t_k^{(a)}z^k\right)
\exp\left(\mp \sum_{k=1}^\infty \frac{\partial}{\partial  t_k^{(a)}}\frac{ z^{-k}}{k}\right).
\end{equation}
For $g_0\in{\cal O}_0\cup\{ 0\}$, where ${\cal O}_0
= R(GL_{\infty})|0\rangle \subset F^{(0)}$ we write: 
\[
\sigma (g_0)= \sum_{{m_1,\ldots, m_s\in\mathbb{Z}}\atop {m_1+\ldots +m_s=0}}\tau^{(m_1,m_2,\ldots,m_s)}(t) q_1^{m_1}q_2^{m_2}\cdots q_s^{m_s}\, \mbox{ and} \quad   \sigma (|0\rangle)=1,
\]
where $t=(t_k^{(a)})_{a=1,\ldots,s,\  k=1,2,\ldots}$.
Then as before 
\begin{equation}
\label{coeftau}\tau^{(m_1,m_2,\ldots,m_s)}(t)=\mbox{ coefficient of }Q_1^{m_1}Q_2^{m_2}\cdots Q_s^{m_s}|0\rangle\mbox{ in }\exp\left(\sum_{a=1}^s\sum_{i=1}^\infty t_i^{(a)}\alpha_i^{(a)} \right) g_k\, .
\end{equation}
Under the isomorphism $\sigma$ we can rewrite (\ref{mKP}) for $k=0$ using (\ref{2m}), to obtain a Hirota bilinear identity for these tau-functions\\
\begin{equation}
\label{mKP2}
\begin{aligned}
&{\rm Res} \, dz \sum_{a=1}^s 
(-1)^{m_1+\ldots+ m_{a-1}+q_1+\ldots +q_{a-1}}
z^{m_a-q_a-2}
\exp\left(\sum_{i=1}^\infty (t_i^{(a)}-y_i^{(a)})z^i\right)\times\\
&
\exp\left(\sum_{i=1}^\infty \frac{\frac{\partial}{\partial y_i^{(a)}}-\frac{\partial}{\partial t_i^{(a)}}}{i}z^{-i}\right)
\tau^{(m_1,\ldots,m_{a-1}m_a-1,m_{a+1},\ldots ,m_s)}(t)\tau^{(q_1,\ldots,q_{a-1},q_a+1,q_{a+1},\ldots ,q_s)}(y) =0.
\end{aligned}
\end{equation}
Note that this equation also holds for $k\not= 0$.
As in the 1-component case we want to describe all polynomial tau-functions of this hierarchy. Of course in this case this is a collection of tau-functions. The approach is similar.
First of all, we let in (\ref{mKP}) $k=m,$ a positive integer, and, as in the
one-component case,
we would like to consider $g_m$ instead of $g_0$, where $m$ is a positive integer
such that
 a polynomial tau-function corresponds to an element 
\begin{equation}
\label{mlambda2}
f_{1}\wedge f_{2}\wedge\cdots\wedge f_{m}\wedge |0\rangle
\end{equation}
where 
\begin{equation}
\label{mlambda3}
f_{j}= \sum_{a=1}^s \sum_{\ell=1}^{M_{j}^{(a)}} b_{\ell j}^{(a)} e_\ell^{(a)},\ j=1, \dots, m,\  \mbox{with } M_{j}^{(a)}\ge 1, \ b_{\ell j}^{(a)}\in\mathbb{C}.
\end{equation}

Note that we may assume this without loss of generality.
Indeed, instead of calculating (\ref{mlambda2}), one could also calculate 
\[
f_{1}\wedge f_{2}\wedge\cdots\wedge f_{m}\wedge Q_1^{r_1}Q_2^{r_2}\cdots Q_s^{r_s}|0\rangle\, .
\]
This gives the same  polynomial tau-function  but translated over the lattice $$(m_1,m_2,\ldots ,m_s)\mapsto (m_1+r_1,m_2+r_2,\ldots ,m_s+r_s)\, ,$$ 
i.e., now 
\[
f_{j}= \sum_{a=1}^s \sum_{\ell=1}^{M_{j}^{(a)}} b_{\ell j}^{(a)} e_{\ell+r_a}^{(a)},\quad \mbox{with }M_j^{(a)}\ge 1,\  b_{\ell j}^{(a)}\in\mathbb{C}\,.
\]
\indent The first step is to determine (cf. (\ref{lambda5}))
\begin{equation}
\label{l3}
\begin{aligned}
f_j(t)=&\exp\left(\sum_{a=1}^s\sum_{i=1}^\infty t_i^{(a)}\alpha_i^{(a)} \right)f_j \\
=&\sum_{a=1}^s 
\sum_{\ell=1}^{M_j^{(a)}}
\sum_{i=0}^\infty
 b_{\ell j}^{(a)} s_i(t^{(a)})e_{\ell-i}^{(a)}
\, .
\end{aligned}
\end{equation}
Note that we can remove all $e_i^{(a)}$ with $i\le 0$ in the above expression, since these elements already appear in $|0\rangle$. Thus
\begin{equation}
\label{l3a}
f_j(t)=\sum_{a=1}^s 
\sum_{\ell=1}^{M_j^{(a)}}
\sum_{i=0}^{M_j^{(a)}-\ell}
 b_{\ell+i, j}^{(a)} s_i(t^{(a)})e_{\ell}^{(a)}\, .
\end{equation}
As before, the coefficient of $e_\ell^{(a)}$, which is equal to 
$\sum_{i=0}^{M_j^{(a)}-\ell}
 b_{\ell+i, j}^{(a)} s_i(t^{(a)})$,  is the $\ell$-th derivative of $\sum_{i=0}^{M_j^{(a)}}
 b_{i, j}^{(a)} s_i(t^{(a)})$ with respect to $t_1^{(a)}$. 
As in the one component case, we can find constants $c_j^{(a)}=(c_{1j}^{(a)},c_{2j}^{(a)},\ldots)$ such that a sum of elementary Schur functions can be expressed as one elementary Schur function with shifted $t$. Thus,
\[
\sum_{i=0}^{M_j^{(a)}}
 b_{i, j}^{(a)} s_i(t^{(a)})=  b_{M_j^{(a)}, j}^{(a)}s_{M_j^{(a)}}(t^{(a)}+c_j^{(a)})
\]
and 
\[
f_j(t)=\sum_{a=1}^s 
b_{M_j^{(a)}, j}^{(a)}
\sum_{\ell=1}^{M_j^{(a)}}
\frac{\partial^\ell s_{M_j^{(a)}}(t^{(a)}+c_j^{(a)})}{\partial (t_1^{(a)})^\ell}
e_{\ell}^{(a)}\, .
\]
Next, define
\begin{equation}
\label{l4}
h_j(t) =\sum_{a=1}^s b_{M_j^{(a)},j}s_{M_j^{(a)}}(t^{(a)}+c_j^{(a)})
\, , 
\end{equation}
then 
\begin{equation}
\label{l5}
f_j(t)=\sum_{a=1}^s\sum_{\ell=1}^{M_j^{(a)}} \frac{\partial^\ell h_j(t)}{\partial (t_1^{(a)})^\ell}\, e_\ell^{(a)}\, .
\end{equation}
Now,
$\tau^{(m_1,m_1,\ldots ,m_s)}(t)$, is the coefficient of
\begin{equation}
\label{l6}
Q_1^{m_1}Q_2^{m_2}\cdots Q_s^{m_s}|0\rangle= e_{m_1}^{(1)}\wedge\cdots\wedge e_{1}^{(1)}\wedge e_{m_2}^{(2)}\wedge\cdots \wedge e_{1}^{(2)}\wedge  e_{m_3}^{(3)}\wedge\cdots e_1^{(s)}\wedge|0\rangle
\end{equation}
	in (\ref{mlambda2}) where $f_j(t)$ is given by (\ref{l5}). It follows from (\ref{mlambda2} and(\ref{mlambda3})  that all $m_a\ge 0$ and $m_1+m_2+\cdots +m_s=m$. Hence, all the labels of non-zero tau-functions lie in the convex polyhedron:
\[
\{
(m_1,m_2,\ldots, m_s)\in\mathbb{Z}^s | \, m_i\ge 0\ \mbox{and } m_1+m_2+\cdots +m_s=m \} \, .
\]
It is straightforward to calculate the coefficient of (\ref{l6}) in (\ref{mlambda2}). As in the 1-component case it is the determinant of a matrix, whose $(m_1+m_2+\cdots+m_{a-1}+i,j)$-the entry, 
with $1\le i\le m_a$,  is the coefficient of $e_{m_a-i+1}^{(a)}$ of $f_j(t)$. By (\ref{l5}) this coefficient  is equal to $\frac{\partial^{m_a-i+1}h_j(t)}{\partial (t_1^{(a)})^{m_a-i+1}}$.
More explicitly , since the group orbit in the one and multi-component case is the same, we only have another realization of the module and hence the orbit, the calculations  prove the following:
\begin{theorem}
\label{mT}
All polynomial tau-functions of the $s$-component KP hierarchy (\ref{mKP2}) are, up to a shift over the lattice, of the form
\[
\tau_m(q,t)=\sum_{m_i\ge 0,\atop m_1+\ldots+m_s=m}\tau^{(m_1,m_1,\ldots ,m_s)}(t)q_1^{m_1}q_2^{m_2}\cdots q_s^{m_s},
\]
where $m$ is a positive integer, and $\tau^{(m_1,m_1,\ldots ,m_s)}(t)$ is given by 
\begin{equation}
\label{ttau}
\tau^{(m_1,m_1,\ldots ,m_s)}(t)=\det\left(\, 
\begin{matrix}
\frac{\partial^{m_1}h_1(t)}{\partial (t_1^{(1)})^{m_1}}&\frac{\partial^{m_1}h_2(t)}{\partial (t_1^{(1)})^{m_1}}&\cdots &\frac{\partial^{m_1}h_m(t)}{\partial (t_1^{(1)})^{m_1}}\\[3mm]
\frac{\partial^{m_1-1}h_1(t)}{\partial (t_1^{(1)})^{m_1-1}}&\frac{\partial^{m_1-1}h_2(t)}{\partial (t_1^{(1)})^{m_1-1}}&\cdots &\frac{\partial^{m_1-1}h_m(t)}{\partial (t_1^{(1)})^{m_1-1}}\\
\vdots&\vdots&&\vdots\\
\frac{\partial h_1(t)}{\partial t_1^{(1)}}&\frac{\partial h_2(t)}{\partial t_1^{(1)}}&\cdots &\frac{\partial h_m(t)}{\partial t_1^{(1)}}\\[3mm]
\hdashline[2pt/2pt]\\[-3mm]
\frac{\partial^{m_2}h_1(t)}{\partial (t_1^{(2)})^{m_2}}&\frac{\partial^{m_2}h_2(t)}{\partial (t_1^{(2)})^{m_2}}&\cdots &\frac{\partial^{m_2}h_m(t)}{\partial (t_1^{(2)})^{m_2}}\\
\vdots&\vdots&&\vdots\\
\frac{\partial h_1(t)}{\partial t_1^{(2)}}&\frac{\partial h_2(t)}{\partial t_1^{(2)}}&\cdots &\frac{\partial h_m(t)}{\partial t_1^{(2)}}\\[3mm]
\hdashline[2pt/2pt]\\[-5mm]
\vdots&\vdots&&\vdots\\
\hdashline[2pt/2pt]\\[-3mm]
\frac{\partial^{m_s}h_1(t)}{\partial (t_1^{(s)})^{m_s}}&\frac{\partial^{m_s}h_2(t)}{\partial (t_1^{(s)})^{m_s}}&\cdots &\frac{\partial^{m_s}h_m(t)}{\partial (t_1^{(s)})^{(m_s)}}\\
\vdots&\vdots&&\vdots\\
\frac{\partial h_1(t)}{\partial t_1^{(s)}}&\frac{\partial h_2(t)}{\partial t_1^{(s)}}&\cdots &\frac{\partial h_m(t)}{\partial t_1^{(s)}}\\
\end{matrix}\, \right).
\end{equation}
Here
\[
h_j(t) =\sum_{a=1}^s b_{M_j^{(a)},j}s_{M_j^{(a)}}(t^{(a)}+c_j^{(a)})
\, , 
\]
where  $M_{j}^{(a)}$ are arbitrary positive integers , $c_j^{(a)}=(c_{1j}^{(a)},c_{2j}^{(a)},c_{3j}^{(a)},\ldots)$  and $b_{M_j^{(a)},j}$ for $1\le j\le m$  and $1\le a\le s$ are  arbitrary constants.
\end{theorem}

\section{The $n$-KdV}
\label{S6}
Let $n$ be an integer, $n\geq 2$. The $n$-th Gelfand-Dickey hierarchy,  or $n$-KdV, describes in the 1-component case the group orbit in a projective representation of the loop group of $SL_n$.
This is not a subgroup of $Gl_\infty$, one has to take a 
bigger group, $\hat{GL}_\infty$ containing it, as, e.g in \cite{KP}. Then the representation $R$ of
$GL_\infty$ extends to a projective representation, denoted by $\hat{R}$, of $\hat{GL}_\infty$.
We obtain the loop algebra $gl_n(\mathbb{C}[x,x^{-1}])$ of  $gl_n$ as follows, see \cite{KP}. Let $\mathbb{C}[x,x^{-1}]$ be the algebra of Laurent polynomials $L$ and denote by $u_1,u_2,\dots,  u_n$ a basis of 
$\mathbb{C}^n$. Identifying 
\[
x^{-k}u_j=e_{nk+j},
\]
we can identify 
\[
x^kE_{ij}=\sum_{\ell\in\mathbb{Z}}E_{\ell n+i, (\ell+k)n +j}\, . 
\]

As explained in \cite{KP}, Lemma 9.1, the centralizer in $\hat{GL}_\infty$ of all operators $\Lambda_n^j= \sum_{\ell\in\mathbb{Z}}E_{\ell,\ell+jn}$, $j\in\mathbb{Z}$, is the central extension of the loop group of $SL_n$ times the scalar operators. Furthermore, the orbit in $F^{(k)}$ of the vacuum vector $|k\rangle$ under the action of the latter group is the intersection of  $GL_\infty |k\rangle$ with the kernels of all operators $\Lambda_n^j$, $j=1,2,\ldots$. Hence all polynomial tau-functions of the $n$-KdV hierarchy correpond to the vectors (\ref{lambda2}) in $F^{(k)}$, satisfying the following condition
\begin{equation}
\label{ncond}
\hat r(\Lambda_n^j)(f_{1}\wedge f_{2}\wedge\cdots\wedge f_{m}\wedge e_{k-m}\wedge e_{k-m-1}\wedge\cdots)=0,\quad j=1,2,\ldots .
\end{equation}
Since, $\hat r(\Lambda_n^j)=\alpha_{jn}$, we find 
that the corresponding polynomial tau-function $\tau_k\in{\cal O}_k$
satisfies $\frac{\partial \tau_k}{\partial t_{jn}}=0$ for $j\ge 1$. Using this, we can differentiate the KP hierarchy  (\ref{KP2}) by $t_{jn}$, which gives the $n$-KdV hierarchy of Hirota bilinear equations on tau-functions:
\begin{equation}
\label{nKdV2}
{\rm Res}\,  z^{jn} \tau_k(t-[z^{-1}])\tau_k(y+[z^{-1}]) \exp\left(\sum_{i=1}^\infty (t_i-y_i)z^i\right)dz=0, \quad j=0,1,2,\ldots\, .
\end{equation}

Let $\hat G$ be the central extension of the loop group of $SL_n$ times the scalar operators. Then equation (\ref{nKdV2}) indeed describes the $\hat G$-group orbit of $|k\rangle$ for non-zero tau-functions. 
Note that $\sigma(\hat R(A)|k+jn\rangle)=\tau_k(t)q^{k+jn}=q^{jn} \sigma(\hat R(A)|k\rangle)$, $j\in \mathbb{Z}$, and  $A\in \hat G$, since $A=(a_{ij})$ has the property that $a_{i+n,j+n}=a_{ij}$. Thus, see e.g. \cite{KP}, Theorem 5.1,  $\tau_k$ indeed satisfies (\ref{nKdV2}).

Next,
(\ref{nKdV2}) implies that 
\begin{equation}
\label{nKdV22}
{\rm Res}\,  \frac{\partial \tau_k(t-[z^{-1}])}{\partial t_{jn}}\tau_k(y+[z^{-1}]) \exp\left(\sum_{i=1}^\infty (t_i-y_i)z^i\right)dz=0, \quad j=0,1,2,\ldots\, .
\end{equation}
Now divide this equation by $ \frac{\partial \tau_k(t)}{\partial t_{jn}}\tau_k(y)$ and let 
\[
	Q(t,z)=\frac{1}{\frac{\partial \tau_k(t)}{\partial t_{jn}}}\frac{\partial \tau_k(t-[z^{-1}])}{\partial t_{jn}},\quad P(t,z)=\frac{\tau_k(t-[z^{-1}])}{\tau_k(t)}.
\]
Then, using pseudo-differential operators, see e.g. \cite{KL} or \cite{KvdL-MKP}, equation (\ref{nKdV22})  implies $(Q(t,\partial)\circ P(t,\partial)^{-1})_-=0$, which  gives that $Q(t,\partial)\circ P(t,\partial)^{-1}=1$. Thus $P(t,\partial)=Q(t,\partial)$, from which one can deduce that $ \frac{\partial \tau_k(t)}{\partial t_{jn}}$ is proportional to $\tau_k(t)$.
 Since we assume that $\tau_k$ is a polynomial, this scalar must be 0. Hence,  $\sigma^{-1}(\tau_k(t)q^k)$ is in the intersection of  $GL_\infty |k\rangle$ with the kernels of all operators $\Lambda_n^j$, $j=1,2,\ldots$. Therefore $\sigma^{-1}(\tau_k(t)q^k)$ is in the $\hat G$-orbit of $|k\rangle$.

Now equation (\ref{nKdV2}) for $j=1$ is the $n$-th modified KP equation. We assume that   $\tau_k(y)$, the second tau-function in (\ref{nKdV2}), lies to the $k$-th charge sector and corresponds to 
\[
F_k=f_{1}\wedge f_{2}\wedge\cdots\wedge f_{m}\wedge e_{k-m}\wedge e_{k-m-1}\wedge\cdots\, ,
\]
where we  may assume that the  $f_j$'s are a linear combination of $e_i$'s with $i>k-m$. Then  $\tau_k(t)$, the first tau-function in (\ref{nKdV2}), lies in the $(k+n)$-th charge sector (recall $j=1$) and thus corresponds to
\[
F_{k+n} =\Lambda_n^{-1} f_{1}\wedge \Lambda_n^{-1} f_{2}\wedge\cdots\wedge \Lambda_n^{-1} f_{m}\wedge e_{k-m+n}\wedge e_{k-m+n-1}\wedge\cdots\, .
\]
Moreover, in the Grassmannian picture, this $n$-modified KP equation, means that the linear span of the factors in $F_k$  is a linear subspace, of codimension $n$, of the  span of the factors in $F_{k+n}$ \ \cite{KP}. Thus every $f_j$  is a linear combination of 
$\Lambda_n^{-1} f_{1}$,  $\Lambda_n^{-1} f_{2}$, $\ldots$, $\Lambda_n^{-1} f_{m}$, $e_{k-m+n}$, $e_{k-m+n-1}$\, $\ldots$.
Thus $\Lambda_n f_j$ is a linear combination of 
  $f_1$, $f_2$,$ \ldots$,  $f_m$, $e_{k-m}$, $e_{k-m -1}$, $\ldots$,  $e_{k-m-n+1}$. Hence, one can choose  $g_i$,  $1\le i\le r<s$  of the form
\begin{equation}
\label{gj}
g_j=e_{\mu_j}+\sum _{i=k-m+1}^{\mu_j-1}c_{ij}e_i
\end{equation}
such that
\begin{equation}
\label{fg}
\begin{aligned}
f_{1}&\wedge f_{2}\wedge\cdots\wedge f_{m}\wedge e_{k-m}\wedge e_{k-m-1}\wedge\cdots=\\
&c\, 
g_1\wedge \Lambda_n g_1\wedge\cdots\wedge \Lambda_n^{\lceil \frac{\mu_1-k+m-n}n\rceil}g_1\wedge g_2\wedge\cdots\wedge 
\Lambda_n^{\lceil \frac{\mu_2-k+m-n}n\rceil}g_2\wedge g_3\wedge\ldots \\
&\qquad \ldots\wedge \Lambda_n^{\lceil \frac{\mu_r-k+m-n}n\rceil}g_r\wedge e_{k-m}\wedge e_{k-m-1}\wedge\cdots\, ,
\end{aligned}
\end{equation}
for certain $0\ne c\in\mathbb{C}$. Here  $\lceil x\rceil$ stands for the ceiling of $x$, i.e. the smallest integer greater or equal to $x$. 
One finds these $g_i$ as follows. In this construction we assume that $m$ is minimal. 
Choose $g_1=f_1$, then all $\Lambda_n^p g_1$, $p\ge 0$,  are in the span of the factors in $F_k$. Moreover $ \Lambda_n^{\lceil \frac{\mu_1-k+m-n}n\rceil }g_1$ still contains vectors $e_i$ with $i>m-k$, but $ \Lambda_n^{\lceil \frac{\mu_1-k+m-n}n\rceil +1}g_1$ is expressed in the $e_i$ with $i\le m-k$, hence  it lies  in the span of the factors in 
$|m-k\rangle$ and $ \Lambda_n^{\lceil \frac{\mu_1-k+m-n}n\rceil }g_1$ does not. Next, choose from the vectors $f_2, f_3, \ldots , f_m$ the first one, say $f_{i2}$ that is not in the span of $\Lambda_n^p g_1$, with $0\le p\le \lceil \frac{\mu_1-k+m-n}n\rceil$, and set
$g_2=f_{i2}$. Again all $\Lambda_n^p g_2$, $p\ge 0$,  are in the span of the factors in $F_k$ and $ \Lambda_n^{\lceil \frac{\mu_{2}-k+m-n}n\rceil +1}g_2$ is in the  span of the factors in 
$|m-k\rangle$. Then choose $g_3=f_{i_3}$, where $f_{i_3}$ is the first $f$ that is not in the  span of $\Lambda_n^p g_a$, with $0\le p\le \lceil \frac{\mu_a-k+m-n}n\rceil$, and $a=1,2$, and continue in this way. After $r<n$ choices this stops. 

This makes it possible to obtain all polynomial tau-functions for the $n$-KdV hierarchy. 
To be more precise, in \cite{KvdL-MKP} we introduced to each partition $\lambda=(\lambda_1,\lambda_2,\ldots,\lambda_m)$ the set
\[
V_\lambda=\{ \lambda_1,\lambda_2-1,\lambda_3-2,\ldots,\lambda_m-m+1,-m,-m-1,-m-2,\ldots\},
\]
and gave the following
\begin{definition}
  A partition $\lambda$ is called $n$-periodic if the corresponding infinite sequence $V_\lambda$
  is mapped to itself when subtracting $n$ from each term.
\end{definition}
This reflects the condition (\ref{ncond}).
Thus we obtained the following \cite{KvdL-MKP}:
\begin{theorem}
\label{TKdV}
All polynomial tau-functions of the $n$-KdV hierarchy are, up to a constant factor, of the form
\begin{equation}
\label{PtKdV}
\tau^n_{\lambda_1,\lambda_2,\ldots,\lambda_k}(t;c_{\overline{\lambda_1}},c_{\overline{\lambda_2-1}},\ldots ,c_{\overline{\lambda_k-k+1}})=\det \left( s_{\lambda_i+j-i}(t_1+c_{1,\overline{\lambda_i-i+1}},t_2+c_{2,\overline{\lambda_i-i+1}}\ldots)\right)_{1\le i,j\le k}\, ,
\end{equation}
where $\overline{i}\equiv i \mod n$ and $\lambda=(\lambda_1,\lambda_2, \ldots,\lambda_k)$ is an $n$-periodic partition. Here  the $c_{\overline i}=(c_{1\overline i},c_{2\overline i},\dots)$ for $i=1,2,\ldots n$ (where at most $n-1$ of such  $\overline i$'s appear)
are arbitrary constants.
\end{theorem}

\section{The $(n_1,n_2,\ldots,n_s)$-KdV}
\label{S8}
In this section we want to consider a reduction of the $s$-component KP hierarchy, which describes again the loop group orbit of $SL_n$, where $n=n_1+n_2+\cdots+n_s$, with 
$n_1\ge  n_2\ge\ldots\ge n_s\ge 1$. 
The case $s=1$ is the $n$-th Gelfand-Dickey hierarchy. The case $n=s=2$, i.e. $n_1=n_2=1$, is the  AKNS (or non-linear Schr\" odinger) hierarchy.

From now on let 
\[
n=n_1+n_2+\cdots+n_s, \quad\mbox{where } n_1\ge  n_2\ge\ldots\ge n_s\ge 1
\]
The identification with $gl_n(\mathbb{C}[x,x^{-1}])$ is via 
\[
x^{-k}u_j^{(a)}=e_{n_ak+j}^{(a)}.
\]
Here the $u_j^{(a)}$, with $1\le a\le s$ and $1\le j\le n_a$, is a basis of $\mathbb{C}^n$.
In this case, it is convenient to relabel the one-component basis in a periodic way to obtain the $s$-component one, i.e., if $e_j^{(a)}=e_i$, then $e_{i+n}=e_{j+n_a}^{(a)}$.
As explained  in Section \ref{S6}, an  element of the central extension of the loop group of $SL_n$ commutes with all the elements
\[
\Lambda^j:=\Lambda_n^j=\sum_{a=1}^s \Lambda_{n_a}^{(a)j}\, ,
\]
 where 
\[
 \Lambda_{n_a}^{(a)j}e_\ell^{(b)}=\delta_{ab}e_{\ell-jn_a}^{(a)}\quad\mbox{and }\hat r(\Lambda_{n_a}^{(a)\,j})=\alpha^{(a)}_{jn_a}, \ \mbox{for }j\ge 1\, .
\]
Note that 
\begin{equation}
\label{Dj}
\sigma\Lambda^j\sigma^{-1}=D_j:=\sum_{a=1}^s\frac{\partial }{\partial t^{(a)}_{jn_a}},\quad j=1,2,\ldots \, .
\end{equation}
This means that the collection of tau-functions $\tau^{(m_1,m_1,\ldots ,m_s)}$ of the $s$-component KP hierarchy satisfies the conditions
\begin{equation}
\label{ref1}
D_j( \tau^{(m_1,m_2,\ldots ,m_s)}(t))=\sum_{a=1}^s\frac{\partial \tau^{(m_1,m_2,\ldots ,m_s)}(t)}{\partial t^{(a)}_{jn_a}}=0,\quad\mbox{for }j=1,2,\ldots \, .
\end{equation}
Thus, letting $D_j$ (only in $t$ not in $y$) act on  equation  (\ref{mKP2}) gives zero on the tau-functions, but acting on the exponetial it produces in every component  a power of $z^{jn_a}$. 
Hence the 
$(n_1,n_2,\ldots,n_s)$-KdV
hierarchy is given by the following equations:
\begin{equation}
\label{mn-KdV2}
\begin{aligned}
&{\rm Res} \, dz \sum_{a=1}^s 
(-1)^{m_1+\ldots+ m_{a-1}+q_1+\ldots +q_{a-1}}
z^{m_a-q_a+jn_a-2}
\exp\left(\sum_{i=1}^\infty (t_i^{(a)}-y_i^{(a)})z^i\right)\times\\
&
\exp\left(\sum_{i=1}^\infty \frac{\frac{\partial}{\partial y_i^{(a)}}-\frac{\partial}{\partial t_i^{(a)}}}{i}z^{-i}\right)
\tau^{(m_1,\ldots,m_{a-1}m_a-1,m_{a+1},\ldots ,m_s)}(t)\tau^{(q_1,\ldots,q_{a-1},q_a+1,q_{a+1},\ldots ,q_s)}(y)  =0,\\
&\qquad\qquad\qquad\qquad\qquad\qquad  j=0,1,2,\ldots\, .
\end{aligned}
\end{equation}
In a similar way as in the 1-component case,  see Section \ref{S6}, one can deduce from equation (\ref{mn-KdV2}),  for $j\not=0$, that  (\ref{ref1}) holds for  polynomial tau-functions.

It is now straightforward to construct all polynomial solutions of this hierarchy. Without loss of generality we may assume again that $m=k$ in (\ref{fg}).
As the $g_j$ in (\ref{gj}), we choose $r<n$ linearly independent functions  $g_j$ ($1\le j\le r$) of the form (\ref{mlambda3}).  Let 
\begin{equation}
\label{k_j}
k_j=\mbox{max}\{ \lceil\frac{M_j^{(a)}}{n_a}\rceil -1|\, 1\le a\le s \}, \quad j=1,2,\ldots,r.
\end{equation}
These $k_j$ are determined by the property  
\[
\Lambda^{k_j}g_j\wedge |0\rangle\not = 0,\quad \mbox{and } \Lambda^{k_j+1}g_j\wedge |0\rangle=0\, .
\]
Then a polynomial tau-function corresponds to (cf. the right-hand side  of (\ref{fg}))
\begin{equation}
\label{ptau}
g_{1}\wedge \Lambda g_{1}\wedge\cdots\wedge \Lambda^{k_1}g_1\wedge  g_{2}\wedge\cdots\wedge \Lambda^{k_2}g_2\wedge g_3\wedge \cdots\wedge \Lambda^{k_r} g_{r}\wedge |0\rangle\, .
\end{equation}
Note that $r+\sum_{i=1}^r k_i=m$.
Define $h_j(t)$ (related to such a $g_j$) again  by (\ref{l4}),
then
$\tau^{(m_1,m_1,\ldots ,m_s)}(t)$ is still given by (\ref{ttau}), but with $h_1(t),h_2(t),h_3(t), \ldots, h_m(t)$ replaced by (cf. (\ref{fg}) and  (\ref{Dj}))
\[
h_1(t), D_1h_1(t),D_2h_1(t),\ldots, D_{k_1}h_1(t), h_2(t), D_1h_2(t), \ldots D_{k_2}h_2(t), h_3(t),\ldots, D_{k_r}h_r(t)\, .
\] 
More specifically, 
an element in the $SL_n$-loop group orbit, where $n=n_1+n_2+\cdots+n_s$ corresponds to a semi-infinte wedge  (\ref{ptau}). 
Then (cf. (\ref{l3})) 
\[
\begin{aligned}
\exp\left(\sum_{a=1}^s\sum_{i=1}^\infty t_i^{(a)}\alpha_i^{(a)} \right)(\Lambda^pg_{j})=&
\sum_{a=1}^s 
\sum_{\ell=1}^{M_j^{(a)}}
\sum_{i=0}^\infty
 b_{\ell j}^{(a)} s_i(t^{(a)})\Lambda^p e_{\ell-i}^{(a)}\\
=&\sum_{a=1}^s 
\sum_{\ell=1}^{M_j^{(a)}}
\sum_{i=0}^\infty
 b_{\ell j}^{(a)} s_i(t^{(a)})e_{\ell-i-pn_a}^{(a)}\\
=&\sum_{a=1}^s 
\sum_{\ell=1}^{M_j^{(a)}}
\sum_{i=-pn_a}^\infty
 b_{\ell j}^{(a)} D_p(s_{i+pn_a}(t^{(a)}))e_{\ell-i-pn_a}^{(a)}\\
=&D^p(h_j(t))\, .
\end{aligned}
\]
Note that in the above calculations we may replace $D_p$ by $D^p$, where $D=D_1$, but this is only because this operator acts on linear combinations of elementary Schur functions in one set of variables $t^{(a)}$, with $a$ fixed.
Thus,
$\tau^{(m_1,m_1,\ldots ,m_s)}(t)$ is the coefficient of (\ref{l6}) in
\begin{equation}
\label{ref2}
\begin{aligned}
\exp\left(\sum_{a=1}^s\sum_{i=1}^\infty t_i^{(a)}\alpha_i^{(a)} \right)g_{1}\wedge \Lambda g_{1}\wedge \ldots \wedge \Lambda^{k_1} g_{1}\wedge  g_{2} \wedge \cdots\wedge  \Lambda^{k_r} g_{r}\wedge|0\rangle=\\
h_{1}(t)\wedge D h_{1}(t)\wedge \ldots \wedge D^{k_1} h_{1}(t)\wedge  h_{2}(t) \wedge \cdots\wedge  D^{k_r} h_{r}(t)\wedge|0\rangle\, .
 \end{aligned}
\end{equation}
Again, as in the $s$-component KP case, 
$\tau^{(m_1,m_1,\ldots ,m_s)}(t)$, is the coefficient of
\[
Q_1^{m_1}Q_2^{m_2}\cdots Q_s^{m_s}|0\rangle\, 
\]
of expression (\ref{ref2}).
In the same way as in the standard $s$-component KP case, one can calculate this coefficient, which is the determinant of a certain matrix. We thus obtain:
\begin{theorem}
\label{mmT}
All polynomial tau-functions of the $(n_1,n_2,\ldots,n_s)$-KdV hierarchy (\ref{mn-KdV2}) are, up to a shift over the lattice, of the form
\[
\tau(q,t)=\sum_{m_i\ge 0,\atop m_1+\ldots+m_s=r+k_1+\cdots+k_r}\tau^{(m_1,m_1,\ldots ,m_s)}(t)q_1^{m_1}q_2^{m_2}\cdots q_s^{m_s},
\]
where $\tau^{(m_1,m_1,\ldots ,m_s)}(t)=$
\[
\left|\, 
\begin{matrix}
\frac{\partial^{m_1}h_1(t)}{\partial (t_1^{(1)})^{m_1}}&\frac{\partial^{m_1}Dh_1(t)}{\partial (t_1^{(1)})^{m_1}}& \cdots&\frac{\partial^{m_1}D^{k_1}h_1(t)}{\partial (t_1^{(1)})^{m_1}}&
\frac{\partial^{m_1}h_2(t)}{\partial (t_1^{(1)})^{m_1}}&\frac{\partial^{m_1}Dh_2(t)}{\partial (t_1^{(1)})^{m_1}}&\cdots &\frac{\partial^{m_1}D^{k_r}h_r(t)}{\partial (t_1^{(1)})^{m_1}}\\[3mm]
\frac{\partial^{m_1-1}h_1(t)}{\partial (t_1^{(1)})^{m_1-1}}&\frac{\partial^{m_1-1}Dh_1(t)}{\partial (t_1^{(1)})^{m_1-1}}&\cdots&\frac{\partial^{m_1-1}D^{k_1}h_1(t)}{\partial (t_1^{(1)})^{m_1-1}}&
\frac{\partial^{m_1-1}h_2(t)}{\partial (t_1^{(1)})^{m_1-1}}&\frac{\partial^{m_1-1}Dh_2(t)}{\partial (t_1^{(1)})^{m_1-1}}&\cdots &\frac{\partial^{m_1-1}D^{k_r}h_r(t)}{\partial (t_1^{(1)})^{m_1-1}}\\
\vdots&\vdots&&\vdots&
\vdots&\vdots&&\vdots\\
\frac{\partial h_1(t)}{\partial t_1^{(1)}}&\frac{\partial Dh_1(t)}{\partial t_1^{(1)}}&\cdots&\frac{\partial D^{k_1}h_1(t)}{\partial t_1^{(1)}}&
\frac{\partial h_2(t)}{\partial t_1^{(1)}}&\frac{\partial Dh_2(t)}{\partial t_1^{(1)}}&\cdots &\frac{\partial D^{k_r}h_r(t)}{\partial t_1^{(1)}}\\[3mm]
\hdashline[2pt/2pt]\\[-3mm]
\frac{\partial^{m_2}h_1(t)}{\partial (t_1^{(2)})^{m_2}}&\frac{\partial^{m_2}Dh_1(t)}{\partial (t_1^{(2)})^{m_2}}&\cdots&\frac{\partial^{m_2}D^{k_1}h_1(t)}{\partial (t_1^{(2)})^{m_2}}&
\frac{\partial^{m_2}h_2(t)}{\partial (t_1^{(2)})^{m_2}}&\frac{\partial^{m_2}Dh_2(t)}{\partial (t_1^{(2)})^{m_2}}&\cdots &\frac{\partial^{m_2}D^{k_r}h_r(t)}{\partial (t_1^{(2)})^{m_2}}\\
\vdots&\vdots&&\vdots&
\vdots&\vdots&&\vdots\\
\frac{\partial h_1(t)}{\partial t_1^{(2)}}&\frac{\partial Dh_1(t)}{\partial t_1^{(2)}}&\cdots&\frac{\partial D^{k_1}h_1(t)}{\partial t_1^{(2)}}&
\frac{\partial h_2(t)}{\partial t_1^{(2)}}&\frac{\partial Dh_2(t)}{\partial t_1^{(2)}}&\cdots &\frac{\partial D^{k_r}h_r(t)}{\partial t_1^{(2)}}\\[3mm]
\hdashline[2pt/2pt]\\[-5mm]
\vdots&\vdots&&\vdots&
\vdots&\vdots&&\vdots\\
\hdashline[2pt/2pt]\\[-3mm]
\frac{\partial^{m_s}h_1(t)}{\partial (t_1^{(s)})^{m_s}}&\frac{\partial^{m_s}Dh_1(t)}{\partial (t_1^{(s)})^{m_s}}&\cdots&\frac{\partial^{m_s}D^{k_1}h_1(t)}{\partial (t_1^{(s)})^{m_s}}&
\frac{\partial^{m_s}h_2(t)}{\partial (t_1^{(s)})^{m_s}}&\frac{\partial^{m_s}Dh_2(t)}{\partial (t_1^{(s)})^{m_s}}&\cdots &\frac{\partial^{m_s}D^{k_r}h_r(t)}{\partial (t_1^{(s)})^{(m_s)}}\\
\vdots&\vdots&&\vdots&
\vdots&\vdots&&\vdots\\
\frac{\partial h_1(t)}{\partial t_1^{(s)}}&\frac{\partial Dh_1(t)}{\partial t_1^{(s)}}&\cdots&\frac{\partial D^{k_1}h_1(t)}{\partial t_1^{(s)}}&
\frac{\partial h_2(t)}{\partial t_1^{(s)}}&\frac{\partial Dh_2(t)}{\partial t_1^{(s)}}&\cdots &\frac{\partial D^{k_r}h_r(t)}{\partial t_1^{(s)}}\\
\end{matrix}\,\; \right|\, ,
\]
where $D=D_1$ is given by  (\ref{Dj}) and
\[
h_j(t) =\sum_{a=1}^s b_{M_j^{(a)},j}s_{M_j^{(a)}}(t^{(a)}+c_j^{(a)}), \ j=1,\ldots, r\, .
\,  
\] 
Here  $M_{j}^{(a)}$ are arbitrary positive integers, $c_j^{(a)}=(c_{1j}^{(a)},c_{2j}^{(a)},c_{3j}^{(a)},\ldots)$ ,  $b_{M_j^{(a)},j}$ for $1\le j\le r$  and $1\le a\le s$ are  arbitrary constants and 
$k_j$ are non-negative integers defined by (\ref{k_j}).
\end{theorem}

\section{The AKNS hierarchy}
The $(1,1)$-KdV hierarchy is the famous AKNS hierarchy. Making the change of variables 
\[
x_i=\frac12(t_i^{(1)}-t_i^{(2)}),\quad y_i=\frac12(t_i^{(1)}+t_i^{(2)}),
\]
then the tau-functions are independent of all $y_i$.
Setting 
\[
q=-\frac{\tau^{(1,-1)}}{\tau^{(0,0)}},\quad r=\frac{\tau^{(-1,1)}}{\tau^{(0,0)}},\quad  x=2x_1,\quad t=-4ix_2
\]
one obtains the Ablowitz--Kaup--Newel--Segur~(AKNS) system  \cite{AKNS},
\[
i\frac{\partial q}{\partial t}=-\frac12 \frac{\partial^2 q}{\partial x^2}-q^2r,\quad
i\frac{\partial r}{\partial t}=-\frac12 \frac{\partial^2 r}{\partial x^2}+r^2q,
\]
as one of the simplest equations in the hierarchy (see e.g. \cite{KL} for more details).
In this case there is only one  function (\ref{l4}), viz. 
\[
h(t)=h_1(t)=b_1s_{M_1}(t^{(1)}+c^{(1)})+b_2s_{M_2}(t^{(2)}+c^{(2)})\quad\mbox{and } \frac{\partial^k D^jh(t)}{\partial (t_1^{(a)})^k}=b_as_{M_a-k-j}(t^{(a)}+c^{(a)})\, .
\]
Expressing this in the variables $x_i$, one has
\[
\frac{\partial^k D^jh(t)}{\partial (t_1^{(a)})^k}=b_as_{M_a-k-j}(-(-1)^a x+c^{(a)}))\, .
\]
Now let $K=k_1+1$, where $k_1$ is as in Theorem \ref{mmT}, then all non-zero tau-functions $\tau^{(p,K-p)}(x)$, where  $p$ is an  integer, $0\le p\le K$, and 
\[
\begin{aligned}
&\tau^{(p,K-p)}(x)=b_1^pb_2^{K-p}\times\\[2mm]
&\qquad \left|\, 
\begin{matrix}
s_{M_1-1}(x+c^{(1)})&s_{M_1-2}(x+c^{(1)})&\cdots &s_{M_1-K}(x+c^{(1)}) \\
s_{M_1-2}(x+c^{(1)})&s_{M_1-3}(x+c^{(1)})&\cdots &s_{M_1-K-1}(x+c^{(1)}) \\
\vdots&\vdots&&\vdots\\
%
s_{M_1-p}(x+c^{(1)})&s_{M_1-p-1}(x+c^{(1)})&\cdots &s_{M_1-p-K+1}(x+c^{(1)}) \\[2mm]
\hdashline[2pt/2pt]\\[-3mm]
s_{M_2-1}(-x+c^{(2)})&s_{M_2-2}(-x+c^{(2)})&\cdots &s_{M_2-K}(-x+c^{(2)}) \\
s_{M_2-2}(-x+c^{(2)})&s_{M_2-3}(-x+c^{(2)})&\cdots &s_{M_2-K-1}(-x+c^{(2)}) \\
\vdots&\vdots&&\vdots\\
s_{M_2-K+p}(-x+c^{(2)})&s_{M_2-K+p-1}(-x+c^{(2)})&\cdots &s_{M_2-2K+p+1}(-x+c^{(2)}) \\
\end{matrix}\,\; \right|.
\end{aligned}
\]
Here $x$ stands for $x=(x_1,x_2,\ldots)$. Note that $\tau^{(p,K-p)}(x)=0$, if $K>\max (M_1,M_2)$.

\end{document}